\documentclass[11pt,a4paper]{article}
\usepackage{amsmath, amssymb, graphics}
\usepackage{amsthm}
\usepackage{graphicx}
\usepackage{geometry}
\usepackage{url}
\usepackage[linktocpage=false,colorlinks,pdfborder={0 0 0},citecolor=black,urlcolor=black,pdfstartview={FitH},linkcolor=black]{hyperref}
\usepackage[T1]{fontenc}
\usepackage[square,numbers, sort&compress, comma, merge]{natbib} 
\usepackage{amssymb}
\usepackage{amsthm}

\newcommand{\ket}[1]{|#1\rangle}
\newcommand{\bra}[1]{\langle #1|}
\newcommand{\braket}[2]{\langle #1|#2\rangle}
\newcommand{\ev}[1]{\langle #1 \rangle}

\theoremstyle{plain}

\theoremstyle{definition}

\theoremstyle{remark}
\newtheorem{rem}{Remark}

\title{Jensen-Shannon Divergence and Non-linear Quantum Dynamics}

\author{Saeid Molladavoudi\thanks{Laboratory of Computational Sciences and Mathematical Physics, Institute for Mathematical Research, Universiti Putra Malaysia, 43400 UPM Serdang, Selangor, Malaysia, E-mail address: \href{mailto:saeid.molladavoudi@gmail.com}{\texttt{saeid.molladavoudi@gmail.com}}} \and Hishamuddin Zainuddin\footnotemark[1] \footnotemark[2] \and Chan Kar Tim\thanks{Department of Physics, Faculty of Science, Universiti Putra Malaysia, 43400 UPM Serdang, Selangor, Malaysia}
}

\date{\vspace{-5ex}}

\begin{document}

\maketitle

\begin{abstract}
Using the statistical inference method, a non-relativistic, spinless, non-linear quantum dynamical equation is derived with the Fisher information metric substituted by the Jensen-Shannon distance information. Among all possible implications, it is shown that the non-linear Schr\"odinger equation preserves the symplectic structure of the complex Hilbert space, hence a Hamiltonian dynamics. The canonically projected dynamics is obtained on the corresponding projective Hilbert space of pure state density operators.
\end{abstract}

\section{Introduction} \label{sec:intro}

The time-evolution equation of quantum states, namely the \textit{linear} Schr\"odinger equation can be obtained by using the \textit{maximum uncertainty principle} introduced by Jaynes \cite{jaynes1957}, and using the statistical inference method \cite{reginatto1998,frieden1998}. The idea, in essence, is to minimize the appropriate action functional, containing the statistical mechanical action and a term proportional to the Fisher information measure \cite{fisher1925}, and combine the two resulting real equations to one linear complex equation by introducing the wavefunction.

Indeed, minimization of the Fisher information metric, to obtain the linear quantum theory, is subject to set of physical constraints. According to the Cram\'er-Rao inequality in statistical estimation theory \cite{rao1945}, the variance (mean-square error) of any unbiased underlying parameter, $\theta$, of a probability distribution, $\rho(x;\theta)$, is maximized once the Fisher information measure is minimized, hence, the maximum entropy principle.

The axiomatic approach in seeking generalizations beyond the linear theory is studied by Parwani \cite{parwani2004}. In \cite{parwani2005}, he proposed a non-linear version of the Schr\"odinger equation based on the same information theoretic discussions by replacing the Fisher information measure with the relative entropy, or the \textit{Kullback-Leibler information measure} \cite{kullback1951,*kullback1959}, in the appropriate action functional, which is partially motivated from the fact that for two close statistical distributions and to the lowest order, the latter measure is proportional to the former. 

Applying the variational method would then lead to the new quantum Hamilton-Jacobi equation and the continuity equation, which again can be combined, by introduction of the wavefunction, to form the non-linear Schr\"odinger equation. The non-linear term is a function of an invariant length scale which inspires the breaking of space-time symmetries, namely the rotational and Lorentz invariances in small scales, i.e. high energies.

In the current paper, we aim at an optimized version of the non-relativistic, spinless, non-linear quantum dynamical equation, introduced in \cite{parwani2005}, which will be obtained by minimization of the bounded, smooth and always well-defined version of the relative entropy, called the \textit{Jensen-Shannon distance measure} \cite{lin1991}. The common properties as well as the different features in comparison with the Kullback-Leibler non-linear quantum dynamics will be studied. Moreover, the implications this rather information-geometrically motivated non-linear quantum dynamics may have on the structures of the linear theory and the dynamical properties on the complex Hilbert space will be investigated. 

The outline of the paper is as follows. In section \ref{sec:qmandit}, both Kullback-Leibler and Jensen-Shannon measures as well as the statistical inference method are reviewed and the metric property of the latter measure is highlighted in further details. In section \ref{sec:gqd}, by using the statistical inference method, the non-linear quantum dynamics is obtained and the extension to many particles in higher dimensions and also its parametric versions are derived. In section \ref{sec:dynprop}, dynamical properties of the non-linear Schr\"odinger equation are studied, in the language of geometric quantum mechanics, followed by a brief discussion about the metric structure. Finally in section \ref{sec:conc} we summarize the results.

\section{Information Measures and Statistical Inference Method} \label{sec:qmandit}
\subsection{Kullback-Leibler Information Measure} \label{subsec:klim}
The Kullback-Leibler (KL) \textit{divergence}, or information measure, was first introduced in \cite{kullback1951,*kullback1959} as the amount of information for discrimination between two statistical distributions $\rho_0(x)$ and $\rho_1(x)$. It is defined as
\begin{equation}
I_{KL}(\rho_1,\rho_0) =  \int{\rho_1(x) \ln \left(\frac{\rho_1(x)}{\rho_0(x)} \right) \,d^mx \,dt} ,
\label{kleq}
\end{equation}
with $\rho_0(x)$ as a reference probability distribution encoding a prior information about the ensemble. The probability distributions are the functions of collective continuous coordinates $x \equiv x^k$, with $k = 1, 2, \dots m$. 

As it is evident from equation \eqref{kleq}, KL information measure is ``\textit{asymmetric}" with respect to two distributions, or in other words, $I_{KL}(\rho_0(x),\rho_1(x))$ is not the same as $I_{KL}(\rho_1(x),\rho_0(x))$. Therefore, the KL measure cannot be considered as a distance along the curve connecting the points $\rho_0$ and $\rho_1$ on the space of statistical distributions.

Furthermore, the functional form of the equation \eqref{kleq} is not invariant under an arbitrary re-parametrization (except affine re-parametrizations) of the integral \eqref{kleq}, which is in contrast with the characteristics of the geodesics in Riemannian geometry being the invariant distance minimizing the curves under arbitrary re-parametrizations \cite{dabak1992}. 

Different symmetric versions of the KL divergence measure exist \cite{johnson2001}, according to various ways of taking the ``average" of two KL measures, such as Jeffreys-Kullback-Leibler or J-divergence which was introduced in \cite{jeffreys1946,jeffreys1948} as
\begin{equation}
J(\rho_0,\rho_1)= \frac{1}{2} \left[ I_{KL}(\rho_0,\rho_1) + I_{KL}(\rho_1,\rho_0)\right] .
\label{j-divergenceeq}
\end{equation}
It can be considered as the average distance between two probability distributions $\rho_0$ and $\rho_1$. The J-divergence is symmetric and also invariant under the re-parametrization but it doesn't satisfy the `triangle' inequality, and is therefore not considered as the metric distance in the space of statistical distributions. Other problems with the J-divergence are its `unboundedness' and `instability' in practical computations \cite{nielsen2010}.

The following identity holds for the $I_{KL}$,
\begin{equation}
I_{KL}(\rho_0,\rho_1) + I_{KL}(\rho_1,\rho_0) = \int_{\theta=0}^{\theta=1}{I_F(\theta) \, d\theta}
\label{klidentityeq}
\end{equation}
where $\rho(x,\theta=0) = \rho_0(x)$ and $\rho(x,\theta=1) = \rho_1(x)$, and the Fisher information measure, $I_F$, is defined as \cite{fisher1925}
\begin{equation}
I_F(\rho) = \int{\frac{1}{\rho} \left( \frac{d \rho}{dx} \right)^2 \, dx} = \int{\rho \left( \frac{\partial \ln \rho}{\partial x} \right)^2 \,dx}.
\label{fishereq}
\end{equation}
In other words, the integral of the Fisher information measure along the geodesic $\rho_{\theta}$ connecting $\rho_0$ to $\rho_1$ is identified with the J-divergence. 

Choosing $\rho_1(x^k) = \rho(x^k + \Delta x^k)$, the KL measure and J-divergence can be expanded, and to their lowest orders one can obtain
\begin{equation}
I_{KL} \left(\rho(x^k +\Delta x^k), \rho(x^k) \right) \simeq  - \frac{1}{2} \left( \Delta x^k \right)^2 \, I_F(\rho(x))
\label{klexpansioneq}
\end{equation}
and
\begin{equation}
J \left(\rho(x^k),\rho(x^k +\Delta x^k) \right) = - \left( \Delta x^k \right)^2 \, I_F(\rho(x))
\label{j-divergenceexpansioneq}
\end{equation}
over a Euclidean manifold $\left( M, \delta_{ij} \right)$ and $\delta^{ij}$ as the inverse of the Euclidean metric. 

\subsection{Jensen-Shannon Information Measure} \label{subsec:jsim}

To avoid the problems of Kullback-Leibler relative entropy, as stated in sub-section \ref{subsec:klim}, the \textit{Jensen-Shannon information measure}, $I_{JS}$, is introduced by Lin \cite{lin1991}. For a one-dimensional Euclidean parameter space it is defined as 
\begin{equation}
I_{JS} = \frac{1}{2} \left( I_{K}\left( \rho_0,\rho_1 \right) + I_{K}\left( \rho_1,\rho_0 \right)\right)
\label{jsdeq}
\end{equation}
where each $I_K$ in the RHS of equation \eqref{jsdeq} is defined as
\begin{equation}
I_{K}\left( \rho_0,\rho_1 \right) = \int{ \rho_0 \ln \left( \frac{2 \rho_0}{\rho_0 + \rho_1} \right) \,dx \,dt}
\label{k-divergenceeq}
\end{equation}
and it is called the K-divergence. 

Recalling the Shannon entropy
\begin{equation}
H(\rho)= -\int{\rho(x) \ln(\rho(x)) \,dx} ,
\label{shannoneq2}
\end{equation}
the Jensen-Shannon divergence can be written as
\begin{equation}
I_{JS}(\rho_0,\rho_1)= H(\frac{\rho_0 + \rho_1}{2}) - \frac{H(\rho_0)}{2} - \frac{H(\rho_1)}{2} \geq 0 .
\label{jsshannoneq}
\end{equation}

As it is shown in \cite{lin1991}, the following inequality illustrates the upper limit of the Jensen-Shannon information measure
\begin{equation}
0 \leq I_{JS}\left( \rho_0,\rho_1 \right) \leq \frac{1}{4} J\left( \rho_0,\rho_1 \right) .
\label{jsinequalityeq}
\end{equation}
Similar to the discussion of the previous section, for two probability distribtions of $\rho(x)$ and the shifted $\rho(x + \Delta x)$, in the limit $\Delta x \rightarrow 0$, one can find \cite{majtey2005b}
\begin{equation}
I_{JS} \simeq \frac{1}{8} \left( \Delta x \right)^2 I_{F}.
\label{jsdfishereq}
\end{equation}

A generalized version of the Jensen-Shannon divergence measure, for two probability densities with different weights, is also introduced in \cite{lin1991}. Assuming $0 \leq \pi_1 , \pi_2 \leq 1$ as the weights for the probability densities $\rho_1$ and $\rho_2$, with $\pi_1 + \pi_2 = 1$, the generalization of equation \eqref{jsshannoneq} is introduced as
\begin{eqnarray}
I^{(\pi)}_{JS}(\rho_1,\rho_2) & = & H \left( \pi_1 \rho_1 + \pi_2 \rho_2 \right) \nonumber \\
& & - \pi_1 H \left( \rho_1 \right) - \pi_2 H \left( \rho_2 \right) \geq 0 .
\label{parjsdeq}
\end{eqnarray}
This is called the ``$\pi$-parametric Jensen-Shannon divergence". For $\pi \equiv \pi_1 = \pi_2 = 1/2$, it reduces to the equation \eqref{jsdeq}. In one dimension for simplicity, the $\pi$-parametric K-divergence is then equal to
\begin{equation}
I^{(\pi)}_{K}\left( \rho_1,\rho_2 \right) = \int{ \rho_1 \ln \left( \frac{\rho_1}{  \pi \rho_1 + \left( 1 - \pi \right) \rho_2} \right) \,dx \,dt}
\label{parkdeq}
\end{equation}
and the Jensen-Shannon measure can be written as
\begin{equation}
I^{(\pi)}_{JS}(\rho_1,\rho_2)= \frac{1}{2} \left( I^{(\pi)}_{K}\left( \rho_1,\rho_2 \right) + I^{(\pi)}_{K}\left( \rho_2,\rho_1 \right)\right) .
\label{parjsdeq2}
\end{equation}

In \cite{parwani2005}, the weight $\pi$ appeared as a regularising parameter, which in fact transforms the Kullback-Leibler information measure to the \textit{asymmetric} K-divergence defined in equation \eqref{parkdeq}, but not $I^{(\pi)}_{JS}$. As in equation \eqref{jsdfishereq}, in the limit $\Delta x \rightarrow 0$ and for two close probability densities $\rho(x)$ and $\rho(x + \Delta x)$ with the weights of $\pi$ and $(1 - \pi)$ respectively, we have 
\begin{equation}
I^{(\pi)}_{JS} \left( \rho(x), \rho(x + \Delta x) \right) \simeq \frac{\left( \Delta x \right)^2}{2} \pi \left( 1 - \pi \right) I_F \left( \rho(x) \right) .
\label{parjsdfishereq}
\end{equation}

Therefore, according to \cite{majtey2005a,majtey2005b} and the equation \eqref{jsdeq}, the Jensen-Shannon divergence is the symmetrized, well-defined and bounded version of the KL information measure satisfying the triangle inequality. Recalling the definition of a distant vector space, it turns out that the square root of the JS divergence denotes a metric in the space of statistical probability distributions \cite{endres2003,osterreicher2003}, i.e. the space $(\mathcal{M}, \sqrt{I_{JS}})$, with $\mathcal{M}$ as the set of statistical probability distributions, is a metric space. Using the results of Schoenberg \cite{schoenberg1938}, one can prove that the space of probability distributions, with the metric $D_{JS}(\rho_0,\rho_1) = \sqrt{I_{JS}(\rho_0,\rho_1)}$, can be isometrically embedded in a \textit{real} separable Hilbert space \cite{fuglede2004,briet2009}. The importance of the metric property of the Jensen-Shannon divergence will be discussed in the sub-section \ref{subsec:sim} below.

\subsection{Statistical Inference Method} \label{subsec:sim}

The dynamical equations of an ensemble of classical particles, moving in one dimension $x$ for simplicity, can be obtained by minimizing the following \textit{classical} action functional
\begin{equation}
I_{CL}= \int{\rho\left(\frac{\partial S}{\partial t}+\frac{1}{2m} \left(\frac{\partial S}{\partial x} \right)^2 + V \right) \, dx \, dt}
\label{classicalactioneq}
\end{equation}
with respect to $\rho$ and $S$ as independent variables. The resulting equations are the classical Hamilton-Jacobi equation and the continuity equation. The latter equation implies that the total probability, $\int{\rho(x)\,dx} =c$, is conserved. Therefore, in the space of statistical densities, the subspace of normalized densities, where $c=1$ and $\rho(x) \geq 0$ for all $x \in \mathbb{R}$, are selected during this process. 

To derive quantum dynamical equations, some physical constraints have to be imposed to the above-mentioned procedure. As mentioned in section \ref{sec:intro}, according to the Cram\'er-Rao inequality, the variance (mean-square error) of any unbiased parameter, such as the particle's positions, is maximized if the Fisher information measure \eqref{fishereq} is minimized. The constraint can be imposed by the method of Lagrange multipliers, namely by applying the variational technique to the modified action functional $I_{CL} + \zeta I_F$, as in \cite{reginatto1998}, and the resulting equations are the quantum Hamilton-Jacobi equation and the probability conservation equation. 

Geometrically, representing the space of statistical probability distributions as a manifold, the Fisher information measure defines the Riemmanian metric on it \cite{amari2000}. The \textit{embedding} from the normalized submanifold of the statistical densities to the infinite-dimensional \textit{real} Hilbert space of the \textit{square-integrable} functions, $L^2(\mathbb{R})$, is provided by the map $\rho(x) \rightarrow \sqrt{\rho(x)}$ and the corresponding Riemannian metric on the \textit{embedded unit sphere} of the real Hilbert space is the Fisher-Rao metric, which is proportional to the Fisher information metric in form of the log-likelihood functions \cite{brody1998,brody2009b}. In the statistical inference method to obtain the \textit{linear} quantum dynamics \cite{reginatto1998}, the polar representation of the wavefunctions, i.e. $\psi(x) = \sqrt{\rho} \exp(i S(x)/ \hbar)$, is used as the embedding map and one can show that the metric induced on the Lagrangian submanifold of the unit sphere in the \textit{complex} Hilbert space, $\mathcal{H}$, is in fact the Fisher-Rao metric, i.e. when $dS=0$ or $S(x,t) = const$ \cite{facchi2010}.

A rather physical axiomatic approach in seeking generalizations beyond linear quantum theory, namely substituting the Fisher metric by the Kullback-Leibler information measure in the statistical inference method, is studied in \cite{parwani2004}. Therefore, according to the axioms in \cite{parwani2004} and the discussions of the sub-sections \ref{subsec:klim} and \ref{subsec:jsim}, the Jensen-Shannon metric \eqref{jsdeq} is an optimized measure in seeking generalizations beyond linear quantum theory. The construction of the non-linear quantum dynamics from other distance measures, such as Renyi's and Wooter's divergences, are also discussed in \cite{parwani2005}. 

The significance of the distance or metric property of the Jensen-Shannon divergence is even more highlighted by recalling the notion of \textit{distinguishability} between two pure quantum states in Wootters' paper \cite{wooters1981}, i.e. the angle between the associated rays in the Hilbert space is equivalent to the concept of statistical distance. More precisely, the metric property of the Jensen-Shannon divergence, i.e. $D_{JS} = \sqrt{I_{JS}}$, of the statistical space $\mathcal{M}$ is \textit{vital}, at least in the sense of preservation of the notion of distinguishability of pure states in the embedded submanifold of the Hilbert space, as a K\"ahler space. Moreover, the idea of distinguishability of pure states is required for the geometrical interpretation of the transition probabilities in the measurement process to remain meaningful, as the geodesic distance between two \textit{distinguished} states in the (projective) Hilbert space, in the linear theory.

In the subsequent section, by using the inference method of the \textit{maximum uncertainty principle}, both non-parametric and parametric versions of the non-linear Schr\"odinger equation will be derived.

\section{Generalized Quantum Dynamics} \label{sec:gqd}
\subsection{The Non-linear Schr\"odinger Equation} \label{sec:nlqm}

The discussion of the previous sub-sections \ref{subsec:jsim} and \ref{subsec:sim}, motivates one to replace the Kullback-Leibler divergence by the \textit{Jensen-Shannon} information measure introduced in equation \eqref{jsdeq}, and minimize the action functional $I_{CL} + \zeta I_{JS}$ with respect to the real-valued functions $\rho$ and $S$, as in \cite{parwani2005}. 

Applying the variational method on $I_{CL} + \zeta I_{JS}$, with two shifted probability distributions $\rho_0 = \rho$ and $\rho_1(x)=\rho(x+l) \equiv \rho_l$ and the Lagrange multiplier $\zeta=\hbar^2/ ml^2$, the standard continuity equation and the modified quantum Hamilton-Jacobi equation are derived respectively as follows
\begin{equation}
\partial_t \rho + \partial_x \left( \rho \frac{\partial_x S}{m} \right) = 0
\label{continuityeq}
\end{equation}
\begin{equation}
\partial_t S + \frac{1}{2m} \left( \partial_x S\right)^2 + V(x) + Q_{N} = 0
\label{mqhjeq}
\end{equation}
with the following quantum potential
\begin{equation}
Q_{N} = \frac{\zeta}{2} \left[ \ln \frac{4\rho^2}{(\rho+\rho_l)(\rho_{-l}+\rho)} \right]
\label{nlqpeq}
\end{equation}
and $\rho_{-l}=\rho(x-l)$. Combining the two real equations through the mapping $\psi=\sqrt{\rho} \exp(iS/ \hbar)$, the non-linear complex Schr\"odinger equation is obtained as
\begin{equation}
i \hbar \frac{\partial \psi}{\partial t} = - \frac{\hbar^2}{2m} \frac{\partial^2 \psi}{\partial x^2} + V(x) \psi + N(\rho) \psi
\label{nlschroedingereq}
\end{equation}
where the non-linear term is
\begin{equation}
N(\rho) = \frac{\zeta}{2} \left[ \ln \frac{4\rho^2}{(\rho+\rho_l)(\rho_{-l}+\rho)} \right] + \frac{\hbar^2}{2m \sqrt{\rho}} \frac{\partial^2 \sqrt{\rho}}{\partial x^2} .
\label{nltermeq}
\end{equation}
The equations \eqref{nlqpeq} and \eqref{nltermeq}, are the altered versions of the quantum potential and the non-linearity term found in \cite{parwani2005} respectively. Regarding the non-linear Schr\"odinger equation \eqref{nlschroedingereq} and the modified quantum potential \eqref{nlqpeq}, the following remarks are in order:

\begin{rem}
By expanding the $Q_{N}$, one can easily verify that to the zeroth order of $l$, $Q_{N} \simeq Q$ and so $N(\rho) \simeq 0$, where $Q$ stands for the ordinary quantum potential derived from the linear Schr\"odinger equation, i.e. the last term in the RHS of the equation \eqref{nltermeq}. To the leading order, the expansion of the non-linear term $N(\rho)$ is
\begin{equation}
N[\rho](x) \simeq \frac{\hbar^2 \, l^2}{64 \, m} \left( \frac{2 \rho''^2}{\rho^2} - \frac{4 \rho'^2 \, \rho''}{\rho^3} + \frac{\rho'^4}{\rho^4} \right) + O(l^4)
\label{nlexpansioneq2}
\end{equation}
where $\rho'$ and $\rho''$ denote $\partial_x \rho$ and $\partial^2_x \rho$ respectively. Therefore, the first leading order of expansion is of the order of $l^2$.
\label{rem1}
\end{rem}

\begin{rem}
The \textit{scale} invariance property of the non-linear term \eqref{nltermeq}, namely $N(\gamma^2 \rho) = N(\rho)$, allows the solutions of the equation (\ref{nlschroedingereq}) to be \textit{normalizable}, i.e. if $\psi$ is the solution, then so is the $\gamma \psi$. If $\gamma$ is only a phase factor, then the solutions are globally (with respect to the configuration) \textit{phase} invariant.
\label{rem2}
\end{rem}

\begin{rem}
For the Gaussian wavepackets, namely $\psi \propto \exp(-x^2/ b^2)$, $Q_{N}$ leads to a repulsive quantum force for all $x$ and it is the cause for wavefunction dispersion, as in the ordinary quantum potential.
\label{rem3}
\end{rem}

\begin{rem}
Since the ordinary continuity equation is preserved, one can imply that the quantum probability current $\vec{j}$ is also left unchanged. To show it explicitly, one can write the non-linear Schr\"odinger equation \eqref{nlschroedingereq} and its complex conjugate for the generalized Hamiltonian $\hat{\tilde{H}} = \hat{H}_0 + N(\rho)$ and multiply them by $\psi^*$ and $-\psi$ respectively. Using the fact that the non-linear term $N(\rho)$ is a real-valued scalar functional, the summation of the obtained equations would lead to the standard continuity equation, with the quantum probability current $\vec{j}$ defined as in the linear quantum dynamics. It is then implied that the norm-squared $\braket{\psi}{\psi}$ is conserved, for pure quantum state $\psi \in \mathcal{H}$ of the non-linear Schr\"odinger equation \eqref{nlschroedingereq}.
\label{rem4}
\end{rem}
\subsection{Many Quantum Particles in Higher Dimensions} \label{subsec:many}

The discussion of the previous section \ref{sec:nlqm} can obviously be extended to higher dimensions with $n$ particles evolving on a $d$-dimensional Euclidean physical space as the following. The classical action functional of equation \eqref{classicalactioneq} is modified, as in \cite{parwani2005}, as
\begin{equation}
I_{CL}= \int{\rho\left(\frac{\partial S}{\partial t}+\frac{\delta^{ij}}{2m_{(i)}} \frac{\partial S}{\partial x^i}\frac{\partial S}{\partial x^j}+V \right) \, d^{nd}x \, dt}
\label{classicalactioneq2}\end{equation}
for $n$ particles, with $x \equiv x^i$, $i,j = 1, \dots , nd$ and the index $(i)$ as the smallest integer $\geq i/d$, as in \cite{parwani2005}. In other words, $i = (k-1)d+1, \dots, kd$ refers to the coordinates of the $k$th particle with the mass $m_k$. Treating each generalized coordinate of the $nd$-dimensional configuration space separately, the extended Jensen-Shannon divergence \eqref{jsdeq} for the $i$th degree of freedom, i.e. $I^i_{JS}(\rho,\rho_i)$, is equal to
\begin{equation}
\frac{1}{2} \int{ \left[ \rho \ln \left( \frac{2 \rho}{\rho + \rho_i} \right) + \rho_i \ln \left( \frac{2 \rho_i}{\rho + \rho_i} \right) \right] \,dx^i \,dt}
\label{extendedjsdeq}
\end{equation}
where $\rho_i(x) \equiv \rho(x_1,x_2, \dots , x_i+l_{[i]}, \dots, x_{nd},t)$ and the symbol $l_{[i]}$ defined as $[i] \equiv i \bmod{d}$ with the convention $l_{[0]} = l_d$, due to the fact that the invariant length scales $l_{[i]}$ are defined in the $d$-dimensional Euclidean physical space, but not in the $nd$-dimensional configuration space.

As usual, one can apply the variational technique to the extended action functional $I_{CL} + \zeta I_{JS}$, with the latter term defined as
\begin{equation}
\zeta I_{JS} \cong \sum_{i=1}^{nd}{\zeta_i I^i_{JS}}
\label{sumjsdeq}
\end{equation}
and $\zeta_i=\hbar^2/m_{(i)}l_{[i]}^2$ combines the resulting continuity equation and the quantum Hamilton-Jacobi equation, through $\Psi=\sqrt{\rho} \exp(iS/ \hbar)$, to obtain the non-linear Schr\"odinger equation as
\begin{equation}
i \hbar \partial_t \Psi(x,t) = \left( -\frac{\hbar^2}{2m_{(i)}} \delta^{ij} \partial_i \partial_j + V(x) \right)\Psi + N(\rho)\Psi
\label{manyschroedingereq}
\end{equation}
with the non-linear term as
\begin{eqnarray}
\lefteqn{N(\rho) = \sum_{i=1}^{nd}{-\frac{\hbar^2}{2m_{(i)}l^2_{[i]}} \ln \left( \frac{(\rho+\rho_i)(\rho+\rho_{-i})}{4 \rho^2} \right)}} \nonumber \\
& & + \frac{\hbar^2 }{8 m_{(i)}} \delta^{ij}\left( \frac{2 \partial_i \partial_j \rho}{\rho}  - \frac{\partial_i \rho \partial_j \rho}{\rho^2} \right)
\label{manynleq}
\end{eqnarray}
and $\rho_{\pm i}(x) \equiv \rho(x_1,x_2, \dotsc , x_i \pm l_{[i]}, \dots, x_{nd},t)$. 

The non-linear Schr\"odinger equation \eqref{manyschroedingereq} for $n$ non-interacting particles decouples into $n$ independent non-linear Schr\"odinger equations for each individual particle. In other words, when $V \left( \vec{r}_1, \vec{r}_2, \dots , \vec{r}_n \right) = V^{(1)}(\vec{r}_1) \oplus V^{(2)}(\vec{r}_2)  \oplus \dotsb \oplus V^{(n)}(\vec{r}_n)$, the particle `$i$' is not influenced by an action on particle `$j$' by change of $V^{(j)}(\vec{r}_j)$, for either initially entangled or factorized (product) state of the composite system, i.e. $\Psi \left( \vec{r}_1, \vec{r}_2, \dotsc , \vec{r}_n ; t_0 \right) = \psi^{(1)}(\vec{r}_1,t_0) \otimes \psi^{(2)}(\vec{r}_2,t_0) \otimes \dotsb \otimes \psi^{(n)}(\vec{r}_n,t_0)$.

However, this is the direct consequence of the ``\textit{complete separability}" of the non-linear term in equation \eqref{manynleq}, in the sense introduced in \cite{czachor1998}, which is defined as
\begin{equation}
N(\rho) = N_1 \left( \rho^{(1)} \right) \oplus N_2 \left( \rho^{(2)} \right) \oplus \dotsb \oplus N_n \left( \rho^{(n)} \right) ,
\label{separabilityeq}
\end{equation}
where $\rho \left( \vec{r}_1, \vec{r}_2, \dotsc , \vec{r}_n \right) = \rho^{(1)}(\vec{r}_1) \otimes \rho^{(2)}(\vec{r}_2) \otimes \dotsb \otimes \rho^{(n)}(\vec{r}_n)$, with
\begin{equation}
\rho^{(i)}(\vec{r}_i) = \int\dotsi\int \left| \Psi \left( \vec{r}_1, \vec{r}_2, \dotsc , \vec{r}_n ; t \right)\right|^2 \, \prod^{n}_{\substack{j=1 \\ j \neq i}}{d \vec{r}_j},
\label{probabilityeq}
\end{equation}
as the \textit{reduced probability density} of the $i$th particle, in which the integration for that particle is singled out. In this sense, the non-linear Sch\"odinger equation for each particle is written independently as
\begin{equation}
i \hbar \, \partial_t \psi^{(i)} \left( \vec{r}_i , t \right) = \left( H^{(i)}_0 + N_i(\rho^{(i)}) \right) \psi^{(i)} \left( \vec{r}_i , t \right)
\label{individualnlseq}
\end{equation}
where
\begin{equation}
H^{(i)}_0 = -\frac{\hbar^2}{2 \, m_i} \Delta_i + V^{(i)}(\vec{r}_i) .
\label{individualheq}
\end{equation}

As it is discussed in \cite{czachor1998}, a consistent extension from a single particle system to a many-particle composite system is guaranteed by the complete separability of the nonlinear Hamiltonian operator in terms of reduced densities of individual particles, i.e. the equation \eqref{separabilityeq}, which will not lead to the locality problem \cite{jordan1990,polchinski1991,czachor1991}, for entangled or generally mixed state of the composite system. Furthermore, by using the approach introduced in \cite{czachor2002}, this construction can be generalized to the multiple-time correlation experiments on the many-particle composite system, without unphysical nonlocal effects, such as those proposed by Gisin in \cite{gisin1989,*gisin1990}.

\subsection{Parametric Non-linear Schr\"odinger Equation} \label{subsec:parametricnlqm}

Recalling the definition of the $\pi$-parametric Jensen-Shannon information measure in subsection \ref{subsec:jsim} and the discussion of the subsection \ref{sec:nlqm}, one is motivated to repeat the same procedure, but this time with $I^{(\pi)}_{JS}$. For one particle moving in one dimension, adopting the weight $\pi$ for the the probability distribution $\rho(x)$ and so $(1 - \pi)$ for the shifted density $\rho(x + l) \equiv \rho_{l}$, the action functional that has to be minimized is then $I_{CL} + \eta I^{(\pi)}_{JS}$. 

As in the previous sections, the Lagrange multiplier $\eta$ is fixed from comparing equation \eqref{parjsdfishereq} with $\hbar^2/8m$, which is used in obtaining the linear quantum theory \cite{reginatto1998}, and therefore
\begin{equation}
\eta = \frac{\hbar^2}{4 \pi \left( 1 - \pi \right) m l^2} .
\label{parlmeq}
\end{equation}
The variational method with respect to $\rho$ and $S$, leads to the ordinary continuity equation and the \textit{parametrically} modified quantum Hamilton-Jacobi equation with the parametric quantum potential term as the following
\begin{equation}
Q^{(\pi)}_{N} = \frac{(4 \pi)^{-1} \hbar^2}{ \left( 1 - \pi \right) m l^2}  \ln \left[ \frac{\rho \left( \pi \rho + \left( 1 - \pi \right) \rho_l \right)^{-\pi}}{\left( \pi \rho_{-l} + \left( 1 - \pi \right) \rho \right)^{(1 - \pi)}} \right] .
\label{parqpeq}
\end{equation}
Combining the resulting parametric Hamilton-Jacobi and continuity equations, through the identification $\Psi = \sqrt{\rho} \exp(i S/ \hbar)$, leads to the following `parametric' non-linear Schr\"odinger equation, for one particle in one dimension for simplicity, as
\begin{equation}
i \hbar \frac{\partial \psi}{\partial t} = - \frac{\hbar^2}{2m} \frac{\partial^2 \psi}{\partial x^2} + V(x) \psi + N^{(\pi)}(\rho) \psi
\label{parnlschroedingereq}
\end{equation}
with the non-linearity term defined as usual as
\begin{equation}
N^{(\pi)}(\rho) = Q^{(\pi)}_{N} - Q
\label{parnltermeq}
\end{equation}
where the $Q$ is the ordinary quantum potential in the linear quantum dynamics. 

\section{Dynamical Properties} \label{sec:dynprop}
\subsection{Background} \label{subsec:background}
The geometrical settings of the linear quantum mechanics has been studied thoroughly so far, for instance in \cite{kibble1979,cirelli1983,heslot1985,anandan1990,ashtekar1997,brody2001} and \cite{schilling1996,bengtsson2006} just to mention a few in the existing literature. It is well-known that the space of  pure quantum states, namely the complex Hilbert space $\mathcal{H}$, is equipped with the structure of a K\"ahler  manifold, $(\Omega, \, G, \, J)$ with $\Omega$ as the non-degenerate symplectic form, $G$ as the metric structure and $J$ as the \textit{compatible complex} structure, which is a linear operator representing multiplication by $i$ and satisfying $J^2=-1$.

Considering $\mathcal{H}$ as a real vector space, the $\Omega$ and $G$ can be identified with the real and imaginary parts of the Hermitian inner product \cite{ashtekar1997,schilling1996}
\begin{equation}
\braket{\Psi}{\Phi} = \frac{1}{2} G \left( \Psi,\Phi \right) + \frac{i}{2} \Omega \left( \Psi, \Phi \right)
\label{innereq}
\end{equation}
with the following relationship
\begin{equation}
\Omega \left( \Psi, \Phi \right) = G\left( J \Psi,\Phi \right)
\label{smeq}
\end{equation}
where $\braket{\Psi}{J \Phi} = i \braket{\Psi}{\Phi}$. For both finite-dimensional and infinite-dimesional Hilbert space, one can show that the linear Schr\"odinger equation is equivalent to the Hamiltonian dynamics, generated by the expectation value of the associated Hamiltonian operator \cite{kibble1979,cirelli1984,heslot1985,ashtekar1997}, namely
\begin{equation}
d H_0(Y) = \Omega \left( X_{\hat{H}_0} , Y\right)
\label{hamdynhileq}
\end{equation}
for any vector field $Y \in \mathfrak{X(\mathcal{H})}$ and $H_0 = \braket{\Psi}{H_0 \Psi} / \braket{\Psi}{\Psi}$ and 
\begin{equation}
X_{\hat{H}_0} \left( \Psi \right) := - \frac{i}{\hbar} \hat{H}_0 \Psi \, .
\label{schrodingervecfieldeq}
\end{equation}

However, the Hilbert space $\mathcal{H}$ is not the true space of states in quantum mechanics, since any two state vectors $\Psi, \Phi \in \mathcal{H}$, such that $\Psi = \alpha \Phi$ with $\alpha \in \mathbb{C}$, are in fact physically equivalent ($\Psi \sim \Phi$). Therefore, the appropriate quantum space of states is the space of rays through the origin of $\mathcal{H}$
\begin{equation}
\mathcal{P(\mathcal{H})} := \mathcal{H} / \sim ,
\label{bundleeq}
\end{equation}
which is called the ``\textit{complex projective Hilbert space}" and denoted by $\mathcal{P}(\mathcal{H})$ for the infinite-dimensional $\mathcal{H} $\cite{brody2001,ashtekar1995}. In particular, the above-mentioned construction defines a vector bundle, called the ``\textit{complex line bundle}", over $\mathcal{P(\mathcal{H})}$, with $\mathbb{C}$ as a typical fibre, $GL(1,\mathbb{C}) \cong \mathbb{C} -\left\{ 0 \right\} \equiv \mathbb{C}^*$ as the structure group and the following \textit{canonical projection} map $\Pi: \mathcal{H} \rightarrow \mathcal{P(\mathcal{H})}$ as
\begin{equation}
\Pi: \ket{\Psi} \mapsto \hat{\rho}_{\psi} := \ket{\Psi} \bra{\Psi} .
\label{projectioneq}
\end{equation}

The space $\mathcal{P}(\mathcal{H})$ is also called the `\textit{quantum phase space}' which is equipped with the structure of a K\"ahler manifold $(\omega,g,j)$, induced from the corresponding structure in the Hilbert space, with the Riemannian structure, called the \textit{Fubini-Study} metric. Its pull-back, according to the canonical projection map $\Pi$, is defined as \cite{anandan1990}
\begin{equation}
\frac{ds_{FS}^2}{4} := 1 - \left| \braket{\Psi}{\Phi} \right|^2 = \braket{d \Psi}{d \Psi} - \left| \braket{\Psi}{d \Psi} \right|^2
\label{fsmetriceq}
\end{equation}
where the last equality in RHS of equation \eqref{fsmetriceq} holds for $\ket{\Phi} = \ket{\Psi} + \ket{d \Psi}$. 

In fact, the Riemannian structure of the quantum phase space, is responsible for the probabilistic interpretation of quantum mechanics. In other words, the \textit{geodesic} distance between two \textit{distinguished} points of the quantum phase space with respect to the Fubini-Study metric specifies the probability of the transition between two quantum states in a measurement process. As in $\mathcal{H}$, one can show that the Schr\" odinger equation is the Hamiltonian dynamics in quantum phase space, with the additional property of being \textit{Killing} vector field with respect to the Fubini-Study metric. 

A rather minimal approach towards the generalization of quantum mechanics is the notion of \textit{non-linear quantum dynamics}, associated with the extended Hamiltonian, which contains more general state-dependent functionals, while assuming the preservation of both the symplectic structure $\omega$ and the Riemmanian metric structure, i.e. the Fubini-Study metric $g$ of the $\mathcal{P(\mathcal{H})}$. These types of generalized quantum dynamics are introduced by Mielnik \cite{mielnik1974}, Kibble \cite{kibble1979} and Weinberg \cite{weinberg1989,*weinberg1989b}, and are generally called MKW-type \cite{brody2010}. 

However, as it is discussed in \cite{czachor1996}, implementing the Fubini-Study metric \eqref{fsmetriceq} of the quantum phase space, does not guarantee a unique probabilistic interpretation of the Weinberg-type non-linear quantum mechanics \cite{weinberg1989,*weinberg1989b}. The Hilbert space representation of the Jensen-Shannon metric \eqref{jsdeq}, will be further discussed in the subsection \ref{subsec:metric} below.

Accordingly, the non-linear Schr\"odinger time-evolution equation in $\mathcal{H}$ reduces to the non-linear Hamiltonian dynamics on $\mathcal{P(\mathcal{H})}$ by the projection to $\mathcal{P(\mathcal{H})}$ of the Hamiltonian vector field, which is generated by a general functional of the wavefunctions but not necessarily by the expectation value of the linear Hamiltonian operator. Conversely, the Hamiltonian flow on $\mathcal{P(\mathcal{H})}$ can be lifted to $\mathcal{H}$ which is governed by the non-linear Schr\"odinger equation.

\subsection{Hamiltonian Dynamics} \label{subsec:hamdyn}

In this section, we are assuming that the non-linear quantum dynamical equation \eqref{nlschroedingereq} is \textit{given}. Then we show that the non-linear Schr\"odinger equation is in fact the Hamiltonian dynamics in disguise. To do so, we start from the non-linear Schr\"odinger equation \eqref{nlschroedingereq} for one quantum particle moving in one dimension for simplicity, which is written as
\begin{equation}
i \hbar \frac{\partial \Psi}{\partial t} = - \frac{\hbar^2}{2m} \frac{\partial^2 \Psi}{\partial x^2} + V(x) \Psi + N(\rho) \Psi
\label{nlschroedingereq2}
\end{equation}
with the new Hamiltonian $\hat{\tilde{H}} = \hat{H}_0 + N(\rho)$ and the non-linear term as
\begin{eqnarray}
\lefteqn{N(\rho) = Q_{N} - Q =} \nonumber \\
& & \frac{\zeta}{2}\, \ln \left[ \frac{4\rho^2}{(\rho+\rho_l)(\rho_{-l}+\rho)} \right] + \frac{\hbar^2}{2m \sqrt{\rho}} \frac{\partial^2 \sqrt{\rho}}{\partial x^2}
\label{nltermeq2}
\end{eqnarray}
with $\zeta = \hbar^2 / 2ml^2$. Since the probabilitistic interpretation of the function $\rho(x) = \Psi^*(x) \Psi(x)$ is guaranteed by the continuity equation \eqref{continuityeq}, the average functionals of $\hat{\tilde{H}}$, denoted as $\mathbb{E}_{\rho}(\hat{\tilde{H}}) \equiv \tilde{H}$, can easily be obtained by substituting the polar form of the wavefunction in
\begin{equation}
\tilde{H} = \int{\sqrt{\rho} \, e^{-\frac{i}{\hbar}S} \left(\hat{H}_0 + Q_{N}(\rho) - Q(\rho)\right) \sqrt{\rho} \,  e^{\frac{i}{\hbar}S} \, dx}
\label{eveq}
\end{equation}
which leads to
\begin{equation}
\tilde{H} = \int{ \rho \left( \frac{\left( \partial_x S \right)^2}{2m} + V + \frac{\zeta}{2} \ln \left[  \frac{4\rho^2 (\rho+\rho_l)^{-1}}{(\rho_{-l}+\rho)} \right] \right) \, dx} .
\label{eveq2}
\end{equation}

One has to note that the equation \eqref{eveq2} is not an expectation value of a quantum linear operator, but a generalized non-linear functional of the wavefunction derived as the average of the extended Hamiltonian. Therefore, by using the functional derivatives for canonically conjugate \textit{fields} $\rho$  and $S$, we see that the functional $\tilde{H}$ is the generator of the Hamiltonian vector field on $\mathcal{H}$, since
\begin{equation}
\frac{\delta \tilde{H} \left[ \rho \right] (x)}{\delta \rho(x)} = \frac{1}{2m} \left( \partial_x S \right)^2 + V + \frac{\zeta}{2} \ln \left[ \frac{4\rho^2(\rho+\rho_l)^{-1}}{(\rho_{-l}+\rho)} \right] =  -\partial_t S
\label{hameq2}
\end{equation}
which is one of the Hamilton's equations of motion. The last equality in equation \eqref{hameq2} comes from the real part of the polar decomposition of the non-linear Schr\"odinger equation \eqref{nlschroedingereq2}. For the field $S$ one can also follow the same steps and find out that
\begin{equation}
\frac{\delta \tilde{H} \left[ S \right] (x^{\prime})}{\delta S(x)} = - \frac{1}{m} \, \partial_x \left( \rho \, \partial_x S \right) = \partial_t \rho
\label{hameq3}
\end{equation}
which is also the Hamilton's equation of motion. As for the equation \eqref{hameq2}, the last equality in equation \eqref{hameq3} comes from the imaginary part of the polar decomposition of non-linear Schr\"odinger equation. Hence, equations \eqref{hameq2} and \eqref{hameq3} determine the corresponding Hamiltonian vector field on $\mathcal{H}$, i.e.
\begin{equation}
d \tilde{H} \left(.\right)= \Omega \left( X_{\hat{\tilde{H}}}, . \right)
\label{hameq4}
\end{equation}
where $\Omega(\Psi,\Phi):= 2 \Im(\braket{\Psi}{\Phi})$ is the symplectic 2-form of the \textit{linear} complex $\mathcal{H}$ in equation \eqref{hamdynhileq}, and
\begin{equation}
\dot{\Psi} \equiv X_{\tilde{H}} \left( \Psi \right) := - \frac{i}{\hbar} \hat{\tilde{H}} \Psi = - \frac{i}{\hbar} \left( \hat{H}_0 + N(\rho) \right) \Psi
\label{schroedingerianvectorfieldeq}
\end{equation}
with
\begin{equation}
X_{\hat{\tilde{H}}} \left( \Psi \right) \equiv X_{\hat{H}_0} \left( \Psi \right) + X_{N} \left( \Psi \right) .
\label{hamvecfieldeq}
\end{equation}
Therefore, as in the linear quantum dynamics, non-linear Schr\"odinger equation \eqref{nlschroedingereq2} can be considered as an infinite-dimensional Hamiltonian dynamics, generated by the generalized real-valued functional \eqref{eveq2}, in case of pure quantum states.

Recalling the equation \eqref{projectioneq}, one can obtain the projective dynamical equation and realize that it satisfies the following non-linear von Neumann equation
\begin{equation}
\frac{d \hat{\rho}_{\psi}}{dt} = \frac{i}{\hbar} [ \hat{\rho}_{\psi} , \hat{\tilde{H}} ],
\label{vneq}
\end{equation}
for pure state density operators $\hat{\rho}_{\psi}$, according to the $x$-dependence of the non-linear potential \eqref{nltermeq2}. In general, an extension from pure state non-linear Schr\"odinger equation to mixed state density operator formalism is not unique \cite{czachor1996}. From the equation \eqref{vneq}, it is implied that the scalar product for two different solutions of the non-linear Schr\"odinger equation \eqref{nlschroedingereq2}, i.e. the angle between two rays corresponding to two different solutions of the equation \eqref{nlschroedingereq2}, is not conserved. This will lead to the ``mobility phenomenon'', in the terminology used by Mielnik in \cite{mielnik1985}.

Moreover, the following identity 
\begin{equation}
\Psi \mapsto \lambda \Psi \Rightarrow X_{N} \left( \lambda \Psi \right) = \lambda \, X_{N} \left( \Psi \right) \qquad \lambda \in \mathbb{C}
\label{homoeq1}
\end{equation}
shows that the Hamiltonain vector field on $\mathcal{H} - \left\{ 0 \right\}$ is homogeneous of degree one, or the $\mathbb{E}_{\rho} \left( N(\rho) \right)$, defined as the following
\begin{equation}
\mathbb{E}_{\rho} \left( N(\rho) \right) = \int{\rho(x) \, N(\rho) \, dx} ,
\label{evneq}
\end{equation}
is homogeneous of degree two, i.e.
\begin{equation}
\mathbb{E}_{\rho} \left( N( \left| \lambda \right|^2 \rho) \right) = \left| \lambda \right| ^2 \, \mathbb{E}_{\rho} \left( N(\rho) \right) .
\label{homoeq}
\end{equation}
These are the homogeneity conditions \cite{weinberg1989,ashtekar1997}, which imply that the non-linear dynamics in the Hilbert space $\mathcal{H}$ is equal to the non-linear dynamics on the unit sphere $\mathcal{S(\mathcal{H})}$. Therefore, $\mathbb{E}_{\rho} \left( N(\rho) \right)$ is a real-valued functional, in the extent to which it contributes to the $\mathbb{E}_{\rho}(\hat{H}_0) \equiv \ev{\hat{H}_0}$ and generate a Hamiltonian flow on $\mathcal{H}$, and also the corresponding Hamiltonian dynamics, i.e. the equation \eqref{vneq}, on the quantum phase space. 

Other well-known non-linear quantum dynamics in the literature are the so called ``non-linear Schr\"odinger equation" with the non-linear term as $\left| \psi(x) \right|^2$ and the one introduced by Bialynicki-Birula and Mycielski in \cite{bialynicki1976} with the non-linear term as $- b \, \ln (a^n \rho)$. In \cite{doebner1999}, Doebner \textit{et al} classified the non-linear Schr\"odinger equations from the gauge-theoretical point of view based on two assumptions, `locality' and `separability' of the non-linear terms. As it is discussed in section \ref{sec:gqd}, the non-linear term introduced in this paper, although `separable', is manifestly `non-local' and hence is not included in the gauge-theoretic classification.

\subsection{Some Comments on the Metric Structure} \label{subsec:metric}

The \textit{Hilbert space} representaion of the Jensen-Shannon distance \eqref{jsdeq} is studied in \cite{majtey2005b}, and it is discussed that while the analytical expression may be difficult and even impossible to obtain, for two close probability distributions and to the first non-vanishing term of order $dx^2$, the Jensen-Shannon distance coincides with half of the \textit{pull-back} of the Fubini-Study distance in the Hilbert space.

In \cite{brody2008,brody2009}, it is shown that \textit{constrained} evolutions on (non-singular) \textit{submanifolds} of the $\mathcal{P(\mathcal{H})}$ are indeed non-linear quantum dynamics in general, while preserving the Hermitian inner product of the associated Hilbert space. Recalling the discussion of subsection \ref{subsec:sim}, minimizing the Fisher metric in the statistical inference method, through the Lagrange multiplier technique, is equivalent with enforcing the dynamics to take place in a constrained submanifold of the space of normalized parametric statistical densities. Then the identification $\sqrt{\rho(x)} \exp(i S/ \hbar)$ maps the dynamics onto the unit sphere $\mathcal{S(\mathcal{H})}$ in complex Hilbert space.

From geometrical point of view, by minimizing the Jensen-Shannon distance measure in the statistical inference method, in section \ref{sec:gqd}, with the same embedding map as in the linear quantum mechanics, the \textit{constrained} submanifold of the normalized densities from the space of statistical densities, i.e. $\mathcal{M}$, is mapped onto a (\textit{constrained}) submanifold of the infinite-dimensional complex Hilbert space, which contains the solutions of the non-linear Schr\"odinger equation. 

In fact, changing the constraint by minimizing the Jensen-Shannon distance measure, while using the same statistical inference method and the same embedding map, can obviously alter the constrained submanifold in the complex Hilbert space. Considering the non-linear Schr\"odinger equation of subsection \ref{subsec:hamdyn} from the constrained dynamical point of view, with of course different type of dynamical constraints as in the linear theory and the \textit{algebraic} constraints discussed in \cite{brody2008,brody2009}, one can conclude that the motion takes place in a \textit{constrained} embedded submanifold of the complex Hilbert space. In this sense, more rigorous analyses of the embedded submanifold and the Riemannian metric induced on it are required, which are not studied in this paper.

\section{Conclusions} \label{sec:conc}
In this paper a non-linear version of the Schr\"odinger equation is derived, by applying the statistical inference method and using the Jensen-Shannon distance information. It is discussed that among all advantages it may have over the Kullback-Leibler information measure, such as smoothness and boundedness, the distance or metric property is the most significant one quantum mechanically. Therefore, it suggests that the axiomatic approach, discussed in \cite{parwani2004}, for using a more general information measure in the statistical inference method rather than the Fisher metric, to be equipped with the distance property, for the notion of distinguishability of the pure quantum states.

The shared properties of our non-linear term with its predecessor introduced in \cite{parwani2005} can be stated as follows: the scale invariance of non-linear term allows the states to be normalizable and the non-linear term in the Schr\"odinger equation is separable for either factorized or entangled initial states, in a system containing many particles. On the contrary, some of the different features with the Kullback-Leibler non-linear quantum dynamics are as follows: the repulsive quantum force acts on a Gaussian ansatz in all range of $x$ as in the linear theory and in the perturbative regime the lowest order of approximation of the non-linearity term is of order of $O(l^2)$. 

Furthermore, it is shown that the non-linear quantum dynamical equation, studied in this paper, is in fact Hamiltonian in the complex Hilbert space, which is generated by the average functional of the modified quantum Hamiltonian operator as the Hamiltonian functional. In addition, the scale invariance of the non-linear term allows the pure quantum states to be normalizable and the structure of the unit sphere to be preserved. This property and the phase invariance of $\mathbb{E}_{\rho}(N(\rho))$ are encoded in the homogeneity condition \eqref{homoeq}. Moreover, according to the properties of the non-linear term, the canonically projective dynamics is in fact the non-linear von Neumann equation, for pure states density operators.

Therefore, while the norms and symplectic structures of the complex Hilbert space are shown to be preserved by the information theoretic non-linearity introduced in this paper, the projective dynamics implies the preservation of the symplectic structure of the quantum phase space $\mathcal{P(\mathcal{H})}$. Finally, by considering the statistical inference method from the constrained dynamical viewpoint, further investigations about the resulting embedded submanifold in $\mathcal{H}$ and the induced Riemannian metric on it are required.

\section*{Acknowledgements} \label{acknl}
This work is supported by the Malaysian Ministry Of Higher Education (MOHE), Fundamental Research Grant Scheme (FRGS) with Vote No. 5523927. SM would like to thank Dr. Rajesh Parwani for fruitful discussions. 

\bibliographystyle{unsrtnat}
\bibliography{Bib_nonlinearqm-pla}

\end{document}